\documentclass[usenatbib,useAMS]{mn2e}
\usepackage{times}
\usepackage{epsfig}
\usepackage[usenames,dvips]{color}

\begin{document}
\title[Wave optics in light curves of exoplanet microlensing]
{Studying wave optics in the light curves of exoplanet microlensing}

\author [Ahmad Mehrabi and Sohrab Rahvar]
{Ahmad Mehrabi$^{1}$ and Sohrab Rahvar$^{1,2}$ \thanks{rahvar@sharif.edu} \\
$^1$ Department of Physics, Sharif University of Technology, P.O.Box
11365--9161, Tehran, Iran \\
$^2$ Perimeter Institute for Theoretical Physics, 31 Caroline Street
North, Waterloo, Ontario N2L 2Y5, Canada}

%\author []
%{ \thanks{rahvar@sharif.edu} \\
%Department of Physics, Sharif University of Technology,\\
%P.O.Box 11365--9161, Tehran, Iran}

\maketitle

\begin{abstract}

We study the wave optics features of gravitational microlensing by a
binary lens composed of a planet and a parent star. In this system,
the source star near the caustic line produces a pair of images in which they can play the role of secondary sources for the observer.
This optical system is similar to the Young double-slit experiment.
The coherent wave fronts from a source on the lens plane can form
diffraction pattern on the observer plane. This diffraction pattern
has two modes from the close- and wide-pair images. From the
observational point of view, we study the possibility of detecting
this effect through the Square Kilometer Array (SKA) project in the resonance and high magnification
channels of binary lensing. While the red giant sources do not seem
satisfy the spatial coherency condition, during the caustic
crossing, a small part of source traversing the caustic line can
produce coherent pair images. Observations of wave optics effect
in the longer wavelengths accompanied by optical observations of
a microlensing event provide extra information from the parameter
space of the planet. These observations can provide a new basis for
study of exoplanets.

\end{abstract}

\section{Introduction}
Gravitational lensing is caused by the bending of light rays due to the
gravitational effect of a foreground mass. Depending on the
distribution of mass on the lens plane and on the relative distances of
the lens and source from the observer, multiple images or distortion
in the source shape can be formed. In the case of star-star lensing
inside the Milky Way, the separation between images is less than
few milliarcseconds and the images are unresolvable for the ground
based telescopes. This type of gravitational lensing is termed
gravitational microlensing.

Einstein (1936) derived the gravitational lensing equation, but
it was decades until the first gravitational lensing was observed
in 1979. The source of this lensing was a quasar and observations were
performed in radio frequencies \cite{Walsh79}. A few years later
Paczy\'ski proposed studying the MACHO (Massive Astrophysical
Compact Halo Objects) population in the Galactic halo by the method
of gravitational microlensing \cite{pa1986}. His suggestion was to
observe stars in the Large and Small Magellanic Clouds, counting the
number of microlensing events and measuring their transit times
(Einstein crossing time). Based on this observation one can measure
the contribution of MACHOs to the mass of Galactic halo. In addition to
dark matter studies, another interesting astrophysical application
of microlensing was suggested by Moa~\&~Paczynski (1991), namely the use of
gravitational microlensing to aid in the discovery of exoplanets.

Microlensing effect due to single or multiple lenses have been
studied mainly using geometric optics. An important study of
wave optics features of gravitational lensing was undertaken by
Ohanian (1983), who investigated the magnification of a radio point
source when a galaxy acts as a gravitational lens. He showed that
wave optics smoothes singular features of the light curve at the
position of caustic lines. In another work, Jaroszy\'ski and
Paczy\'ski (1995) studied the caustic crossing of Quasar Q2237+0305
by a galaxy composed of individual stars. By studying the
diffraction images of this system, they could put limit on the size
of the quasar. Wave optics observation of gravitational lensing
inside the Milky Way also have astrophysical applications, for example
studying the limb darkening of small sources like white dwarfs
\cite{zabel diff}. Recently, Heyl (2010,2011a,2011b) discussed
the possibility of detecting wave optics signals in microlensing
light curves with a single substellar lens.

In this work our aim is to extend the application of the wave optics
to the conventional method of extra solar planet detection by
gravitational microlensing. Here we assume a binary lens composed of
a lensing star and a planet. The crossing of the caustic lines of this
system by the source star produces high magnification in the light
curve. Moreover, owing to the small separation of the images on the
lens plane, the gravitational lensing system resembles a multiple
slit optical system in the astronomical scales. With a coherent
condition for the wave fronts on the lens plane, the result would be
a diffraction pattern on the observer plane. We study the
applications of this method in both the resonance and high
magnification channels of the exoplanet detection. Observations of
the contrast in the fringes and transit time of the fringes enable
us to break degeneracy between the lens parameters. We also study
the possibility of observing the wave optics features of binary
microlensing using the future Square Kilometer Array (SKA) project.

In section \ref{Ampl1}, we introduce wave optics formalism in
gravitational lensing and calculate the wave optics light curve for
a binary lens system. In section \ref{near} we carry out
semi-analytic calculations of the wave optics feature for
microlensing near the caustic lines and study the temporal and
spatial coherency conditions. We also numerically compute wave
optics light curves and compare them with the results of geometric
optics. In section \ref{det} we discuss the possibility of detecting
microlensing wave optics signals by a binary lens in which one of the
lenses is a planet. Our study uses observation in
radio or micrometer wavelengths and future observations with SKA. We
also discuss possibility of degeneracy breaking between the lens
parameters in the resonance and high-magnification channels of
exoplanet detection. Conclusion and a summary are given in section \ref{conc}.

\section{Wave Optics in gravitational lensing}\label{Ampl1}
In geometrical optics, the locations of the images in terms of
position of the source can be obtained from the lens equation
\begin{equation}\label{lens-eq}
 \mathbf{y}=\mathbf{x}-\alpha(\mathbf{x}),
\end{equation}
where $\mathbf{x}$ and $\mathbf{y}$ are respectively the angular
positions of the image and source normalized to the projected
Einstein angle in each plane and $\mathbf{\alpha}(x)$ is the
deflection angle, which depends on the distribution of matter as
\begin{equation}\label{deflection}
 \mathbf{\alpha}(\mathbf{x})=\frac{1}{\pi}\int
\kappa(\mathbf{x^{\prime}})\frac{\mathbf{x}-\mathbf{x^{\prime}}}{\vert
\mathbf{x}-\mathbf{x^{\prime}}\vert ^{2}}d^{2}x^{\prime},
\end{equation}
where $\kappa(x) = \Sigma(x)/\Sigma_{cr}$,  $\Sigma(x)$ is the surface mass density of the lens,
and $\Sigma_{cr}^{-1} = (4 \pi G
D_{d}D_{ds})/(c^{2}D_{s})$. Here, $D_{s}$, $D_{ds}$, $D_{d}$ are
the source-observer, lens-source and lens-observer distances,
respectively. Another method for deriving lens equation is to use
the Fermat principle. For the stationary points of the Fermat
potential, the position of images in terms of the position of source
is obtained from
\begin{equation}
\label{ferma-lens}
\nabla_\mathbf{x} \phi(\mathbf{x},\mathbf{y})=0,
\end{equation}
where Fermat's potential is given by
\begin{equation}
\label{ferma-dif}
\phi(\mathbf{x},\mathbf{y})=\frac{1}{2}(\mathbf{y}-\mathbf{x})^{2}-\psi(\mathbf{x}),\nonumber \\
\end{equation}
and the deflection angle is
\begin{equation}
\alpha(\mathbf{x})=\nabla_\mathbf{x} \psi(\mathbf{x}).
\end{equation}
We use the determinant of Jacobian of the mapping function from the
source plane to the lens plane. The magnification of a point-like
source can be obtained as
\begin{equation}
\label{Ampli-geo}
\mu(x_i)=\frac{1}{|det
J(x_{i})|}=\frac{1}{|\phi_{11}(x_{i})\phi_{22}(x_{i})-\phi_{12}^{2}(x_{i})|},
\label{jackobian}
\end{equation}
where $x_{i}$ is the location of the $i$th image. For an extended
source, we should calculate the magnification over the source area.
One of the methods of calculating the magnification by an extended
source is Green's theorem.
In this theorem, a two-dimensional integration over
the source reduces to a one-dimensional integration on the boundary of
the images \cite{Mar07}.

From the Huygens principle in wave optics, every point on the
lens plane can be considered as a secondary source. The amplitude of
the electromagnetic wave at each point of the observer plane is
composed of the superposition of the infinitesimal sources on the
lens plane. From the Kirchhoff integral, we can obtain the amplitude
of the electromagnetic wave $F_{\mu\nu}$ if we know the boundary
condition on the lens plane \cite{born}. Multiplying the
superposition of the electromagnetic wave by its complex conjugate
results in the magnification of the wave on the observer plane as
follows \cite{Schn}:
\begin{equation}\label{Ampli-wav}
\mu(\mathbf{y})=\frac{f^{2}}{4\pi^{2}}\vert \int
e^{if\phi(\mathbf{x},\mathbf{y})}d^{2}x\vert ^{2},
\end{equation}
where $f\phi(\mathbf{x},\mathbf{y})$ corresponds to the phase of the
electromagnetic waves emitted from a source, deflected from the lens
plane and received by the observer. Here $f$ is given by $f =
2kR_{s}$ where $k$ is the wave number and $R_{s}=2GM_{d}/c^{2}$ is
the Schwarzschild radius of the lens. This formula have been
obtained for a monochromatic electromagnetic wave.

In order to show the compatibility of the wave optics formalism with that of
geometric optics and  the transition from the wave optics to the
geometric optics formalism, we expand the Fermat potential in the lens plane
using a Taylor series. For a point like source, the expansion around
the image is given by
\begin{eqnarray}\label{expand-ferma}
\phi(\mathbf{x},\mathbf{y})&=&\phi^{(0)}+(\mathbf{x}-\mathbf{x}_{i}).\nabla_x
\phi^{(0)} +
\frac{1}{2}[(x_{1}-x_{1i})^{2}\phi_{11}^{(0)}  \\
&+&(x_{2}-x_{2i})^{2}\phi_{22}^{(0)}+2(x_{1}-x_{1i})(x_{2}-x_{2i})\phi_{21}^{(0)}]+...,\nonumber
\end{eqnarray}
where the superscript $(0)$ represents the Fermat potential at the
position of the image and subscripts represent the derivatives with
respect to two directions on the lens plane. Substitute from
equation (\ref{expand-ferma}) into (\ref{Ampli-wav}), the first term of eq.(\ref{expand-ferma})
after multiplication of $\exp[i\phi(\mathbf{x}_{i},\mathbf{y}_i)]$
by its complex conjugate results in unity. The second term is zero from the
Fermat principle, and finally the third term as non-zero term
results in the magnification in the geometric optics as equation
(\ref{Ampli-geo}). The third and higher orders of derivatives in the
Fermat potential result in the wave optics features in the light
curve.

One of the important issues in the wave optics formalism is that, in reality,
the source is not completely coherent and we need to define a
coherent time scale of $\Delta\tau = \Delta\omega^{-1}$, where
$\Delta\omega$ is the width of the spectrum. The amplitude of a
non-chromatic source on the observer plane is given by
$$V(\mathbf{x},\mathbf{y},\phi) \propto \int g(\omega) e^{i 2 R_s \phi(\mathbf{x},\mathbf{y})\omega} d\omega,$$
where the magnification is given by
\begin{equation}\label{Ampli-wav2}
\mu(\mathbf{y})=\frac{R_s^{2}}{\pi^{2}}\vert \int d^{2}x \int
g(\omega) e^{2i R_s \phi(\mathbf{x},\mathbf{y})\omega} \omega
d\omega \vert ^{2}.
\end{equation}
we set the speed of light to $c = 1$. For a coherent monochromatic
source $g(\omega) = \delta(\omega-\omega_0)$ when substituting this
specific spectrum in equation (\ref{Ampli-wav2}), we can recover
equation (\ref{Ampli-wav}). Assuming a non-zero temperature for a
monochromatic source, the Doppler broadening can change the spectrum
of a Dirac-delta spectrum to a Gaussian distribution.

An important issue regarding the observability of the wave optics effect
is the coherency of light arriving to the observer from different
parts of the lens plane, the so-called temporal coherency \cite{man81}.
In addition we need to have coherency between different parts of an
extended source, termed spatial coherency. Assuming that a
source has zero angular size, in order to examine the temporal coherency
between different images, we need the time delay between the
light rays from the source to the observer and
compare it with the coherent time of the source. The time difference between the two
light rays received by the observer is given by
\begin{equation}\label{tim-diff}
\Delta
t=2R_{s}\left[\phi(\mathbf{x}_{I1},\mathbf{y})-\phi(\mathbf{x}_{I2},\mathbf{y})\right],
\end{equation}
where the source position is fixed and $\mathbf{x}_{I1}$ and
$\mathbf{x}_{I2}$ are the positions of the images. The difference
between the Fermat potentials, $\Delta\phi$, for two distant images is
of the order of unity. On the other hand, during the caustic
crossing, two pairs of close images can be formed with $\Delta\phi$
of the order of $10^{-3}$. Quantifying the time delay for the wide
and close pair of images in the caustic crossing, $\Delta t$ is
given as
\begin{equation}\label{tim-diff1}
\Delta t \sim  1 \times {2R_{s}} \sim 10^{-5}
\frac{M}{M_\odot}~\rm{sec},~~~ \rm{wide~images} \label{far}
\end{equation}
and
\begin{equation}\label{tim-diff2}
\Delta t \sim  0.001\times {2R_{s}} \sim 10^{-8} \frac{M}{M
\odot}~\rm{sec}~~~ \rm{close~images}, \label{close}
\end{equation}
where in the former case images appear around the star-lens and in
the latter case images appear around the planet-lens. Here we take
the mass ratio of the planet to the parent star to be of the order
of $10^{-3}$. We will discuss this in detail in section
({\ref{near}). As the source approaches the caustic line (i.e.
$\Delta t \rightarrow 0$), the time difference between the light
rays become shorter. For a source with a non-zero temperature $T_{s}$,
the coherent time in terms of the bandwidth of the spectrum is given
by $\tau_c \Delta\omega \sim 1$ \cite{mehta}. The dispersion
velocity of the gas is related to the surface temperature of a star
by $\sigma \sim \sqrt{T_{s}}$, on the other hand, the dispersion
velocity is related to the frequency dispersion as $\sigma =
\Delta\omega/\omega$, hence the coherent time relates to the
temperature of the source as \cite{Guenther}:
\begin{equation}\label{coher-time}
\tau_{c} \propto \frac{1}{\sqrt{T_{s}} \nu} \Rightarrow
\tau_{c}=\frac{2.8\times
10^{-4}}{\nu_{(GHz)}}\sqrt{\frac{3000}{T_{s}}}.
\end{equation}
In the double-slit experiment with a point-like source for a
diffraction pattern, the time-difference between the two light rays
received by the observer should not be longer than the coherent
time. Now we can constrain $f = 2kR_{s}$ with a temporal coherency
condition for the wide and close pair of images. Comparing equations
(\ref{far}) and (\ref{close}) with the coherent time in equation
(\ref{coher-time}) results in
\begin{eqnarray}\label{cond-f1}
f_{\rm{wide}}\leq 10^{6}\sqrt{\frac{3000}{T_{s}}},\\
f_{\rm{close}}\leq 10^{9}\sqrt{\frac{3000}{T_{s}}}.
\end{eqnarray}
More details of temporal coherency in the wave optics gravitational
lensing are given in Appendix (\ref{appA}).
\begin{figure}
\begin{center}
\psfig{file=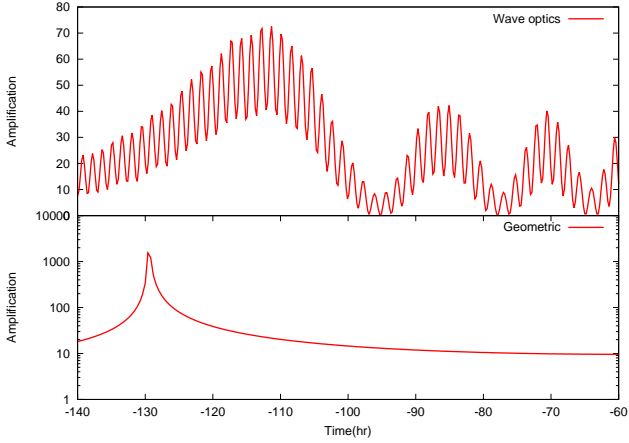,angle=270,width=8.5cm,clip=} \caption{The wave
optics light curve of a point like coherent source (upper panel)
with the parameters of $d=0.8$, $q=0.1$, $u_{0}=0$, $t_{E}=20$~days
and $f=1000$. Lower panel shows the light curve in the geometric
optics. The light curves depict magnification of the source star
during the caustic crossing.} \label{fig1}
\end{center}
\end{figure}

\begin{figure}
\begin{center}
\psfig{file=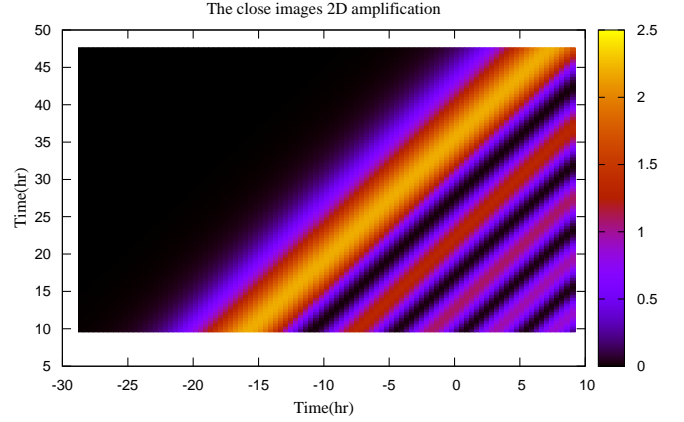,angle=270,width=10.cm,clip=} \caption{The two
dimensional luminosity pattern from a point-like source on the
observer plane lensed by a binary system. The fringes are
demonstrated near a critical line. The overall flux results from the
sum of the close images (as shown in this figure) plus flux from
the incoherent wide images.} \label{fig-2d}
\end{center}
\end{figure}

In order to calculate magnification by a binary lens composed of a
parent star and a planet,  we apply the following Fermat potential
for a binary lens in equation (\ref{Ampli-wav}),
\begin{equation}\label{Fermat}
\phi(\mathbf{x},\mathbf{y})=\frac{1}{2}(\mathbf{x}-\mathbf{y})^{2}-m_{\star}\ln({\vert\mathbf{x}-\mathbf{x}_{\star}}\vert)
-m_{p}\ln(\vert{\mathbf{x}-\mathbf{x}_{p}}\vert),
\end{equation}
where $m_{\star}$ and $m_{p}$ are the relative mass of the star and
planet to the overall mass of the system, respectively. The
positions of the parent star and planet are given by
$\mathbf{x}_{\star}$ and $\mathbf{x}_{p}$. We take lenses along the
$x_{1}$ axis and put the center of mass of this system at
$\mathbf{x}=0$. The positions of parent star and planet are given by
\begin{equation}\label{Ampli-poi}
\mathbf{x}_{\star}=(\frac{d~q}{1+q},0)~~,~~\mathbf{x}_{p}=(-\frac{d}{1+q},0),
\end{equation}
where $d$ is the projected distance between the two lenses and it
has been normalized to the Einstein radius and $q = m_p/m_\star$. We
can identify the track of the source on the lens plane by two
parameters: the minimum impact parameter $u_{0}$ with respect to the
center of mass of lenses and its direction with respect to x-axis,
$\alpha$. For the case of a single lens, the integral in equation
(\ref{Ampli-wav}) has an analytical solution; for two point-mass
lenses, however, we perform numerical computations to obtain the light curve
of a source moving with respect to the lens plane. In the next
section, we utilize the Fourier expansion of the Fermat potential up
to relevant terms, and study the Fermat potential of a binary lens
near the caustic lines.

For a point-like source lensed by a binary system, we use equation
(\ref{Ampli-wav}) and plot the amplification pattern as a function
of time in Fig. (\ref{fig1}). This is a typical microlensing light
curve with the wave optics features compared with those of geometric
optics. Here we have two oscillating modes due to the interference
between the wide and close images. As the source gets closer to the
caustic line, the longer mode is magnified and after caustic
crossing it becomes dimmer. In Fig. (\ref{fig-2d}) we depict the two-dimensional
pattern of fringes on the observer plane when the
relative motion of the observer with respect to this pattern
produces the light curve in Fig. (\ref{fig1}). In the following
section, we include the finite-size effect of the source star in our
calculations.

\subsection{Finite size effect}\label{fin-siz}
The majority of microlensing observations in recent years have been carried out
by two MOA (Microlensing Observation in Astrophysics) and
OGLE (Optical Gravitational Lensing Experiment
observational groups, monitoring millions of
stars toward the Galactic Center. There are other follow-up
telescopes around the world that cover ongoing events for 24 h.
These telescopes with high-cadence observations identify anomalies
in the microlensing light curves to discover extra-solar
planets. While there are all type of stars as microlensing targets
towards the Galactic bulge, there is a selection bias for the red giant
population compare to the main sequence stars \cite{rah12,rahal09}.
In the direction of Galactic bulge, red clump stars contain the majority
of the source stars used for microlensing events
\cite{Hamadache}. This selection of source for the microlensing
events is due to the brightness of red giants, which enable us to observe
them from a distance. Consequently, the averaged Einstein angle
corresponds to this type of source stars is larger than that for main
sequence stars, making the transit time of these events longer. In
addition to the visual band, red giants can optically pump
interstellar medium and produce Maser emissions a few astronomical
units away from the source star \cite{sio maser,water maser}.

In contrast to the simple point-like sources, red giants are
extended objects in which, extended source effect, not only decreases the
strength of the magnification for geometric optics, but also
decreases the enhancements of the fringes for wave optics
\cite{mutual-coh}. The overall magnification for an extended source
is given by \cite{Schn}
\begin{equation}\label{finte}
\mu=\frac{\int_{S}\mu(\mathbf{y}) d^2\mathbf{y}}{\pi \rho^{2}},
\end{equation}
where $\mu(\mathbf{y})$ is the magnification of the multiple images
at the source position of $\mathbf{y}$ and $\rho$ is the angular
radius of source normalized to the Einstein angle. A detailed
expression for $\mu$ is given in equations (\ref{mut}) of Appendix.

The contrast in the interference fringes on the observer plane
depends on the size of source star in such a way that increasing the
source size decreases the contrast of the interference pattern in
the light curve. This effect can be seen in the Young experiment
when we increase the size of the pinhole as the source in double slit
experiment. The superposition of the fringes from different parts of
a source in the observer plane is an indicator for the spatial
coherency of the source. A mathematical criterion for losing spatial
coherency is that the constructive interference from one part of
source overlap with the destructive interference from the other part
of the source in the observer plane.

Now we apply the spatial coherency condition of the Young experiment
to the microlensing effect. Let us consider a source with size $L_s$
located at distance of $D_{ls}$ from the lens plane. For light rays
arriving at the lens plane within a domain of radius $h$ from the
optical axis of the system, the spatial coherency condition is met if
\begin{equation}
L_s<\frac{D_{ls} \lambda}{2h}. \label{diff}
\end{equation}
For a single lens, the angular separation between the images is
given by $\Delta\theta = \sqrt{\beta^2+4\theta_E^2}$, where $\beta$
is the impact parameter. For high-magnification events,
$\Delta\theta \simeq 2 \theta_E$ and the separation between the images
is given by $2h = D_{ol}\Delta\theta \simeq 2 D_{ol} \theta_E$.
Rewriting equation (\ref{diff}) in terms of the source size and
Einstein angle, constraint on the spatial coherency of the source is
given by
\begin{equation}
L_s<\frac{D_{ls}}{D_{ol}}\frac{\lambda}{2\theta_E}.
\end{equation}
Using the definition of the Einstein angle, we have
\begin{equation}
L_s<\frac{\lambda}{2} \sqrt{\frac{D_{ls} D_s}{2R_s D_{ol}}},
\label{constrain}
\end{equation}
where $R_s$ is the Schwarzschild radius of the lens. This
calculation is done for a single lens. In the next section we
recalculate coherency condition for a binary lens, using the Fermat
potential. The advantage of a binary lens is that this lensing
system can produce very close images during the caustic crossing.
These images may satisfy the spatial coherency condition of the
source star. Before studying the spatial coherency of a source in
the binary lensing, let us estimate the coherent size of the source for
a single lens.

We assume a lens at the middle of distance between the observer and
source where the probability of microlensing observation is maximum
(i.e. $D_{ol} = D_{ls}$). For two typical planets with masses of the
Earth and Jupiter, the Schwarzschild radius is about $1$ and
$286$ cm, respectively. For a source star at the Galactic bulge,
$D_s = 8.5$ kpc, Table (\ref{tab1}) shows the coherent size of
sources at various wavelengths. For $\lambda = 3$~cm, we can
observe the wave optics features of an earth-mass lens with a solar
type source star. We can also observe the wave optics effect for a
Jupiter-mass planet and a smaller source. At micron wavelengths,
the spatial coherency decreases to $10-100~km$, where we may detect
diffraction pattern of smaller structures such as granules on the
surface of source star. \cite{yu}.

\begin{table}
\begin{center}
\begin{tabular}{ccc}
\hline
$\lambda$ & {3 cm} & {1 mm}\\
\hline\hline
$L_s(M_\oplus)$  &  $3.4\times 10^6$ & $114$ \\
$L_s(M_J)$  &   $2.0\times 10^5$ & $6.7$ \\
\hline
\end{tabular}
\vspace{1cm} \caption {Coherent size of a source in kilometer for a
single microlensing system. Here $D_s = 8.5 kpc$, $D_l = 4~kpc$. The
observation is performed with the two wavelengths of $3$cm and
$1$mm for the cases of earth and Jupiter mass lenses. Content of
this table shows the size of coherent sources.} \label{tab1}
\end{center}
\end{table}

As already noted, in binary lensing, during the approach of
the source star to a caustic line, the distance between the pair of
images can be very small compare to the case of a single lens. At the same time
we may have wide images located a few astronomical units away from
each other. Hence, the light rays received from the lens plane to
the observer are a mixture of close coherent images and wide
incoherent images. In the next section we discuss the possibility of
producing a diffraction pattern by a binary lens.

\section{Light curve near caustic line: Binary lenses}\label{near}
In this section we use numerical and semi-analytical methods to
study the light curve of a microlensing event by a binary lens.
During the caustic crossing, where images form at the critical
lines, we can write the lens equation. In other word, the first
derivatives of the Fermat potential is zero:
$$\phi^{(0)}_{1}=\phi^{(0)}_{2}=0.$$
Diagonalizing the Fermat potential with respect to the second
derivatives, we can set $\phi^{(0)}_{12} = \phi^{(0)}_{21} = 0$. In
order to satisfy singular Jacobian transformation on the critical
lines, from equation (\ref{Ampli-geo}), either $\phi_{11}$ or
$\phi_{22}$ should be zero. We set $\phi^{(0)}_{22}=0$ and
$\phi^{(0)}_{11}\neq0$. Ignoring $\mathbf{x}^2$ in the geometric
term compare to $\mathbf{y}^2$ term, we perform a Taylor expansion of the
Fermat potential around the critical line as \cite{Schn,pac95}:
\begin{eqnarray}\label{Ferma-appr2}
\phi(\mathbf{x},\mathbf{y})&=&\phi^{(0)}+\frac{1}{2}\mathbf{y}^2-\mathbf{x}.\mathbf{y}+\frac{1}{2}\phi^{(0)}_{11}x_{1}^{2}+
\frac{1}{6}(\phi^{(0)}_{222}x_{2}^{3}  \nonumber \\
& +& \phi^{(0)}_{111}x_{1}^{3}) + \frac{1}{2}(\phi^{(0)}_{112}x_1 + \phi^{(0)}_{122} x_2)x_1 x_2 + ...
\end{eqnarray}
Using the Fermat principle of $\delta\phi/\delta x_i = 0$, we obtain
the position of the images as a function of position of the source,
\begin{eqnarray}
\label{y1}
y_1 &=& \phi^{(0)}_{11} x_1 + \frac12 \phi^{(0)}_{111} x_1^2 + \phi^{(0)}_{112} x_1 x_2, \\
y_2 &=& \frac{1}{2} \phi^{(0)}_{222}x_2^2 + \phi^{(0)}_{122} x_1
x_2. \label{y2}
\end{eqnarray}
Singularity for the new Jacobian of transformation implies the
constrain of $\phi^{(0)}_{122}
x_1 + \phi^{(0)}_{222} x_2 = 0$, where we have ignored the higher order terms of $x$, \\

From equations (\ref{y1}) and (\ref{y2}), we obtain the position of the
images as follows:
\begin{equation}\label{loc-two-imag}
\mathbf{x}_{images}=(\frac{y1}{\phi^{(0)}_{11}},\pm \sqrt{\frac{2y_{2}}{\phi^{(0)}_{222}}}),
\end{equation}
where $y_1$ is chosen along the caustic line and $y_2$ is
perpendicular to the caustic line. On the positive side of $y_2$ we
have two images while for the negative side there is no image.
Substituting the position of the images in equation
(\ref{Ferma-appr2}), we can calculate the Fermat potential for the
nearby images during the caustic crossing. The difference between
the Fermat potentials of the two images is given by
\begin{equation}
\Delta \phi
=\frac{2}{3}\frac{(2y_{2})^{\frac{3}{2}}}{(\phi^{(0)}_{222})^{\frac{1}{2}}}.
\label{dphi}
\end{equation}
Here $\Delta\phi$ is a function of $y_2$ and the third derivative of
Fermat potential on the critical line.  Assuming the trajectory of the
source (i.e. $\vec{y}(t)$) has a direction given by the angle $\gamma$
with respect to the caustic line, the position of the source in the
direction perpendicular to the caustic line
is given by $y_2 = \sin\gamma \times(t -
t_c)/t_E $. On the other hand, from the Fermat potential for a
binary system in equation (\ref{Fermat}), we can calculate
$\phi_{222}^{(cm)}$ in the center of mass coordinate system as
follows:
\begin{eqnarray}\label{phi222}
 \phi_{222}^{(cm)}=m_{p}[\frac{6x_{2}}{((x_{1}-x_{p1})^{2}+x_{2}^{2})^{2}}-\frac{8x_{2}^{3}}{((x_{1}-x_{p1})^{2}+x_{2}^{2})^{3}}]\nonumber \\
+m_{\star}[\frac{6x_{2}}{((x_{1}-x_{\star 1})^{2}+x_{2}^{2})^{2}}-\frac{8x_{2}^{3}}{((x_{1}-x_{\star 1})^{2}+x_{2}^{2})^{3}}].
\end{eqnarray}
Here $x_1$ and $x_2$ represent the position of the critical lines
on which images form during the caustic crossing. The
position of images on the critical line depends on the location of
the source, and double images can form along the critical line,
either around the star ( wide images) or around the planet (close
images). These pairs of images split from single image during the
caustic crossing. Because our calculation has been done on the local
coordinates of the critical line where images form, we perform a coordinate
transformation of the Fermat potential from the center of mass
coordinate system to the local diagonalized coordinate system at the
image position. First we do a boost along the x-axis to one of lens
positions. The second boost is along the radial direction, equal to
the Einstein radius of lens. Finally, we perform a rotation with $R_{ij}$
to diagonalize $\phi_{ij}$. The third derivative of the Fermat
potential which is the relevant parameter in the Fermat potential in
equation (\ref{dphi}) can be obtained after the coordinate
transformation as follows $\phi^{(0)}_{222} =
R_{2i}R_{2j}R_{2k}\phi^{(cm)}_{ijk}$.

The corresponding coordinate transformations to the location of
images around the critical line of the planet is given by $x_1 =
x_{p1} + {R_{E}^{(p)}}/{R_E}\cos\theta$ and $x_2 =
{R_{E}^{(p)}}/{R_E}\sin\theta$ where $R_{E}^{(p)}$ is the Einstein
radius of the planet, $R_E$ is the overall Einstein radius and
$\theta$ is the polar angle with respect to the line connecting the
two lenses. Since $|R_{ij}| \sim 1$, the magnitude of the third
order derivative of the Fermat potential doesn't change so much by
the coordinate transformation, $\phi^{(0)}_{222} \simeq
\phi^{(cm)}_{ijk}$. Substituting boosts in equation (\ref{phi222}),
since $m_\star\gg m_p$, the first term dominates as $ R_E^{(p)}$
appears in the denominator, and hence
\begin{equation}
\label{phi222} \phi_{222}^{(0)} \simeq -m_p
(\frac{R_{E}^{(p)}}{R_E})^{-3}.
\end{equation}
Replacing the ratio of Einstein radius of the planet to the Einstein
radius of the star with the corresponding mass ratio and using a
normalized planet mass with $m_p = q/(1+q)$, equation (\ref{phi222})
can be written as
\begin{equation}
\phi_{222}^{(0)} \simeq -\frac{1}{\sqrt{q}}.
\label{phi222-2}
\end{equation}

Comparing the images that form around the planet and the parent star on
the lens plane, the nearby images around the planet are more
suitable for producing a diffraction pattern on the observer plane.
Having a smaller $q$ results in a larger $\phi^{(0)}_{222}$ and consequently a
smaller $\Delta\phi$. We now substitute equation (\ref{phi222-2}) into
equation (\ref{dphi}) and replace the difference in the Fermat
potential with the difference in the time delay between the
trajectory of the two images as follows:
\begin{eqnarray}\label{del-tim}
\Delta t &=& 2\times{R_{s}}\Delta \phi \nonumber \\
&=& {5 \times10^{-9}}\nonumber \\
&\times& (\frac{t-t_{c}}{1h}\sin\gamma)^{3/2}(\frac{t_{E}}{40days})^{-3/2}(\frac{q}{0.001})^{1/4}(\frac{M}{M\odot})sec
\nonumber
\end{eqnarray}
where $(t-t_c)/1h$ is the time corresponds to the relative distance
of the source from the caustic line which is normalized to one hour.
We note that, unlike for the single lens where $\Delta t$ is of the order of
light crossing time of the Schwarzschild radius, in the case of a
binary lens the factor $q$ decreases the corresponding time-scale
\cite{Heyl10}. On the other hand, as $t\rightarrow t_c$, $\Delta t$
approaches zero. The characteristic time difference in the Fermat
potential is about $\Delta t \sim 5 \times 10^{-9}~s$
corresponding to $\Delta l \sim 150 ~cm$. This length scale
corresponds to the frequency of $0.2~GHz$. Here wavelengths larger
than this threshold $\Delta l$ satisfy temporal coherency and can
produce a diffraction pattern from the images on the observer plane.
Repeating this calculation for the distant images on the lens plane,
where $q$ is of the order of one, we obtain a larger value for $\Delta
t$, destructive for producing the wave optics features. We note that
wavelength $\lambda \sim 150~cm$ is a typical size of wavelength for
temporal coherency and near the caustic crossing as $\Delta l$
approaches zero, we can observe a diffraction pattern in the
shorter wavelengths.

\begin{figure}
\begin{center}
\psfig{file=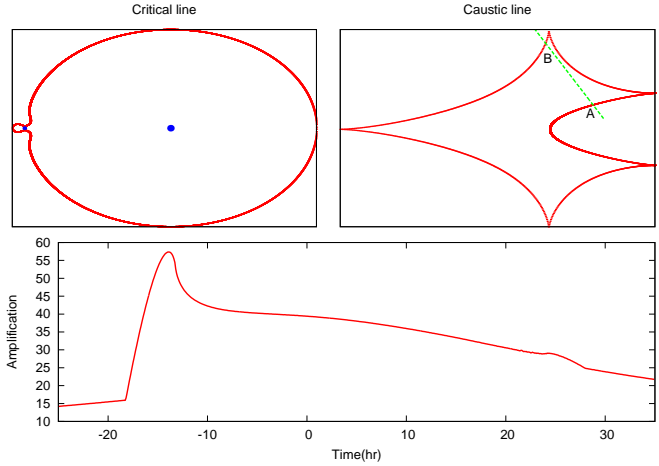,angle=270,width=9.cm,clip=} \caption{Critical
lines (top left panel), caustic lines (top right panel) and light
curve in the geometric optics (lower panel) for a binary system with
the parameters of $q=0.001,~~d =1$. Here we take source star with
the size of $\rho=0.002$. Large loop (at the top left panel)
corresponds to the critical line of the parent star lens and smaller
loop corresponds to the critical line of the planet. The straight
dashed line (at the top right panel) indicates the path of the
source on the source plane.} \label{caus-fig}
\end{center}
\end{figure}

The other essential condition to have wave optics feature, as
discussed in the previous section, is the spatial coherency of the
source. From equation (\ref{loc-two-imag}), the physical distance
between the images on the lens plane, rewriting the dimensionless
parameter of $x_{image}$ in terms of the Einstein radius, is given by
\begin{equation}
h = 2R_E \sqrt{\frac{2y_{2}}{\phi^{(0)}_{222}}}. \label{imagediff}
\end{equation}
Substituting $h$ in equation (\ref{diff}) and expressing $y_2$ in
terms of transit time, the condition of spatial coherency for a
binary lens system is given as follows:
\begin{equation}
L_s<\frac{\lambda D_{ls}}{4 R_E}\sqrt{\frac{t_E~\phi^{(0)}_{222}
}{2(t-t_c)~\sin\gamma}},
\end{equation}
and we can rewrite this equation in terms of characteristic scales
of the lens and source as follows:
\begin{eqnarray}
L_s<1.59\times10^{6}(\frac{\lambda}{10cm})(\frac{D_{ls}}{4kpc})(\frac{R_{E}}{1~a.u})^{-1}\times\nonumber \\
(\frac{t_{E}}{40days})^{0.5}(\frac{q}{0.001})^{-0.25}(\frac{t-t_{c}}{10hr})^{-0.5}~km
. \label{ls}
\end{eqnarray}
Note that $L_s$ on the left-hand side of this inequality represents
the part of the source that crosses the caustic line. Here, only this
part of source contributes in production of close-pair images. Hence
even for larger sources, we may have spatial coherency condition at
the beginning of the caustic crossing.

The other observable parameter which depends on the position of
source and relative motion with respect to the caustic lines is the
magnification. Substituting equation (\ref{Ferma-appr2}) into equation
(\ref{Ampli-wav}) and keeping the leading terms in the Fermat
potential, the magnification near the caustic is obtained as follows
\cite{Schn}:
\begin{equation}\label{Ampli-Airy}
\mu\sim[Ai(\frac{y_{2}}{Y_{0}})]^{2},
\end{equation}
where $Y_{0}=(\frac{\vert\phi_{222}\vert}{2f^{2}})^{\frac{1}{3}}\sim
q^{-1/6}$ and $Ai(x)$ is the Airy function. The maximum
magnification also relates to the derivatives of the Fermat
potential via $\mu_{max} \propto |\phi^{(0)}_{11}|^{-1}
|\phi^{(0)}_{222}|^{-2/3}\sim q^{1/3}$. Near the critical line for
the close-images, larger $\phi_{222}^{(0)}$ gives a smaller maximum
magnification for the fringes in the light curve and larger $Y_{0}$
produces longer modes. On the other hand, for the wide images around
the lens star, $q$ is large, $\phi_{222}^{(0)}$ is small and we have
shorter modes of the diffraction pattern with larger magnification (i.e.
closer to the geometric optics case). We recall that in the diagonalized
coordinate system for the Fermat potential, $\phi_{11}$ is the
non-zero term of the second-order derivative of the Fermat
potential. On the other hand, the trace of $\phi_{ij}$ is
$\nabla^2\phi = 2 - \nabla^2\psi$, where for two point-mass lenses
the second term on the right-hand side of this equation is the Dirac
delta function at the position of the lenses. We can set this
function to zero along the caustic lines, far from the position of
the lenses. Hence for a small distance from the caustic lines,
$\phi_{11} = 2$. Therefore, the only relevant parameters in the
 wave optics light curve are $f$, $\phi_{222}$ and trajectory of the
source with respect to the caustic line.

\begin{figure}
\begin{center}
\psfig{file=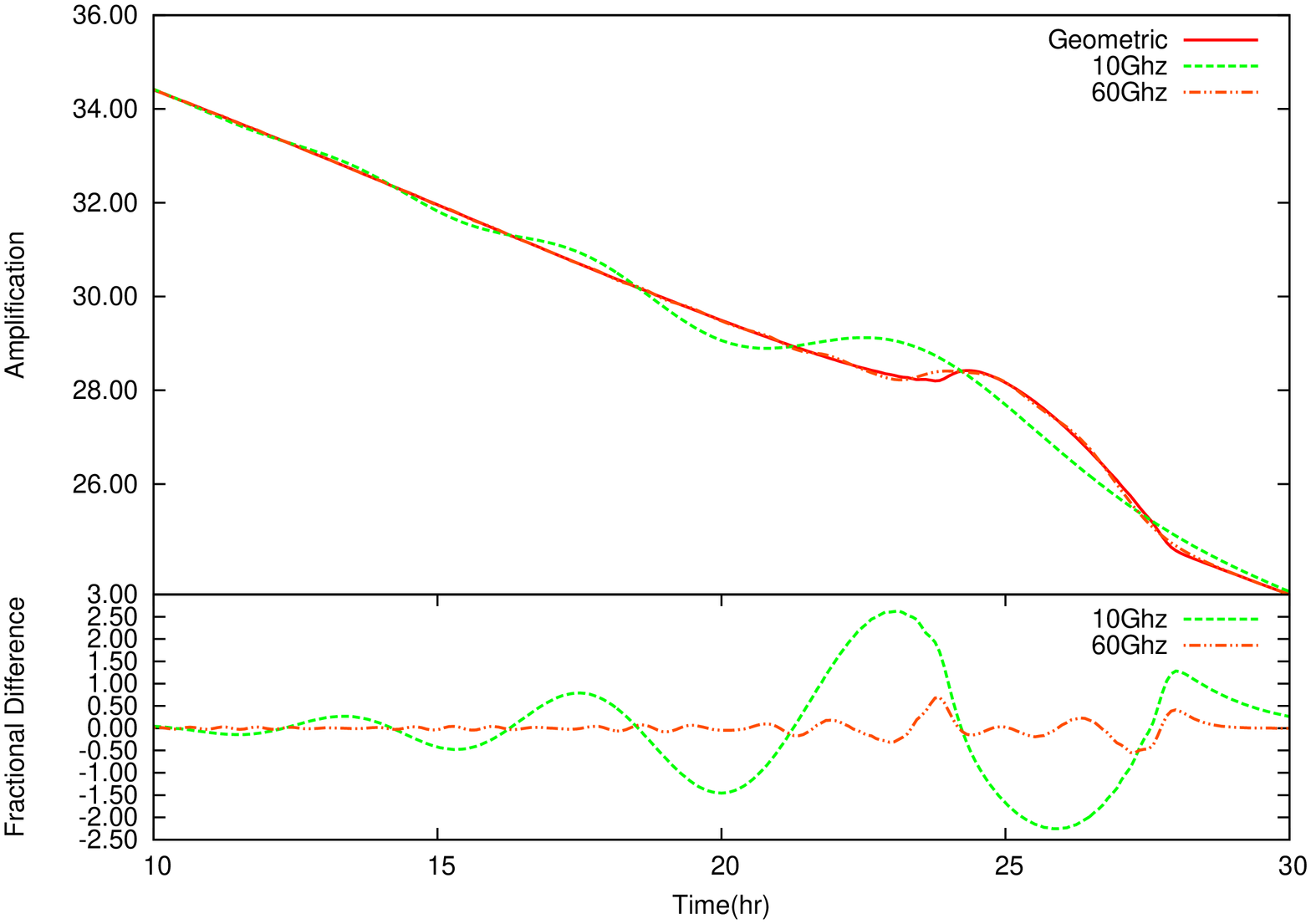,angle=270,width=9.cm,clip=}
\psfig{file=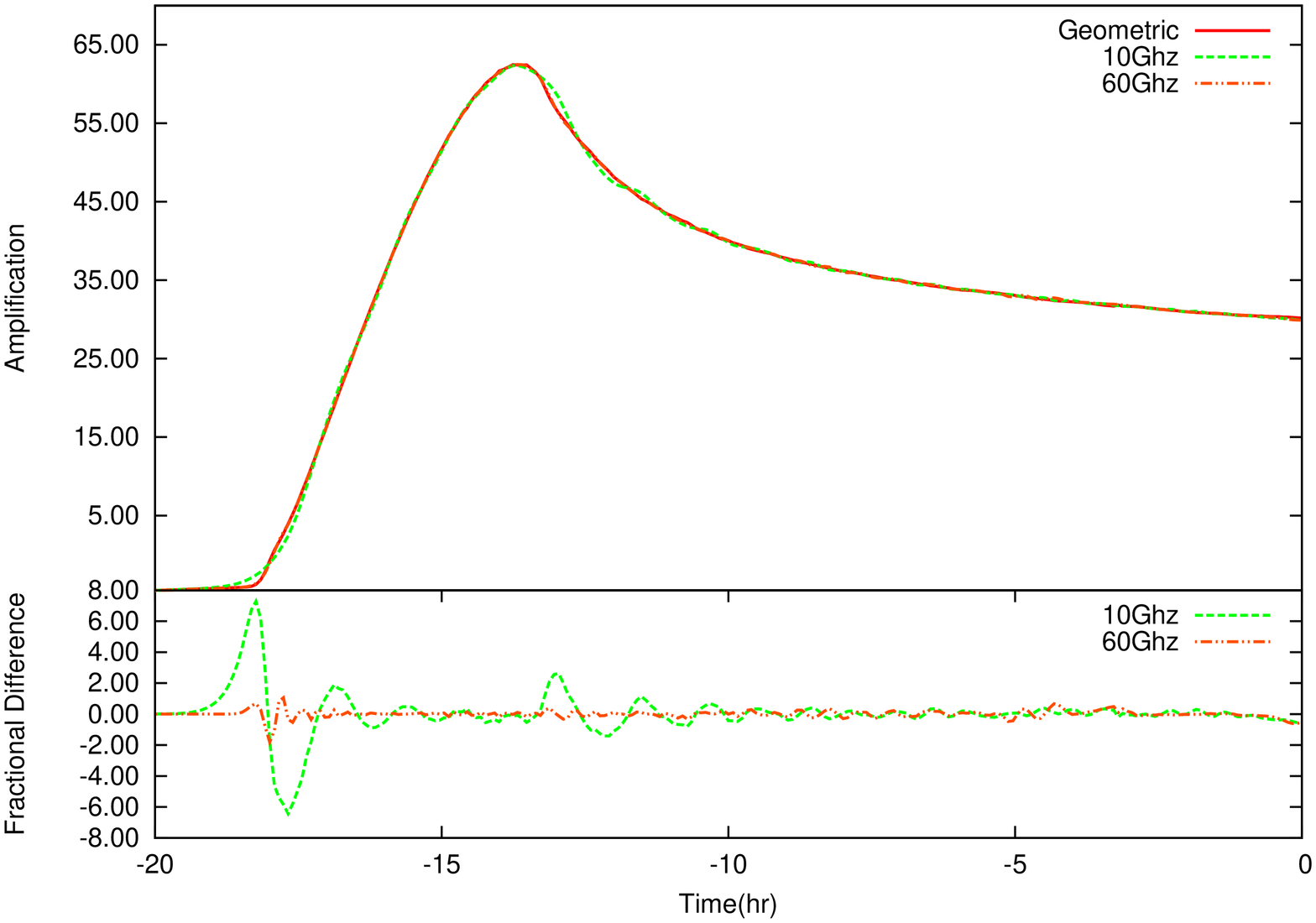,angle=270,width=9.cm,clip=} \caption{The light
curves correspond to the close pair images (upper panel) around the
planet and wide pair images (lower panel) around the parent star in
a binary lenses. Here the source crosses the caustic line at points
$A$ and $B$ (shown in Figure \ref{caus-fig}). The solid line
corresponds to the geometric optics and two wave optics light curves
are shown in $10$ GHz (dashed line) and $60$ GHz (dashed-dotted
line). The smaller panels in each figure correspond to the
fractional difference (in percentage) between the geometric optics
and the wave optics.} \label{caus}
\end{center}
\end{figure}

Fig. (\ref{caus-fig}) shows the configuration of a binary lens (parent
star and planet) with the corresponding critical and caustic lines
and the light curve for geometric optics. Caustic crossing takes
place at points ${A}$ and ${B}$, producing close and wide images.
Fig. (\ref{caus}) also compares the light curves in the wave optics and
 geometric optics formalisms for two wavelengths at points $A$ and $B$. The
light curves are obtained from the Fresnel-Kirchhoff integration.
For the close images, the time variation of the fringes is slower
than for the wide images. This effect can be seen from the relative
velocity between the separation of the pair images from equation
(\ref{imagediff}) as follows:
\begin{equation}
\dot{h} = v_t\sqrt{\frac{2 t_E
}{|t-t_c|}\times\frac{\sin\gamma}{\phi_{222}^{(0)}}}, \label{vrel}
\end{equation}
where $v_t = R_E/t_E$ is the relative velocity of the source with
respect to the lens. For the close-pair images, we have a larger
$\phi_{222}^{(0)}$ and for the wide-pair images this term is smaller.
 Hence not only close-images produce large modes of fringes, their
time variation also slower and provides enough time for the
observer to detect this effect with a suitable cadence of
observations.

\section{Observational Prospect}\label{det}
In this section we study the possibility of follow-up observations
of a binary microlensing system during the caustic crossing by the
future SKA project. For a single lens, the parameters involved in the
light curve are the Einstein crossing time $t_E$, the minimum impact
parameter $u_0$ and the time for the maximum magnification $t_0$.
Amongst these parameters, the only parameter that contains the
physical information of the system is $t_E$ which is a function of the
mass of lens, the relative distance of the lens with respect to the source
and observer and the transverse velocity of the lens with respect to
our line of sight. Using additional information such as the parallax
effect due to the annual motion of the earth around the Sun
\cite{gould}, one can partially break the degeneracy between the
lens parameters \cite{rahvar}. However, the finite-size
effect of the source can also provide extra information to break the
degeneracy between the lens parameters \cite{fsf}.

Increasing the number of lenses from one to two increases the
number of parameters of the lensing system. The additional
parameters are (i) the projected distance between the lenses
normalized to the Einstein radius, $"d"$, (ii) the relative mass of the
lenses, $"q"$ and (iii) the angle $"\alpha"$ defining the trajectory of
the source with respect to the line joining the two lenses. For a
binary system, we have a total of  six parameters to fit the light
curve. The probability of microlensing observation of a binary lens depends on
the size of the caustic and for the case of resonance where the
distance between the lenses is of the order of Einstein ring, we
will have the maximum probability of detection.

The main problem with the resonance events is that, in spite of
occasional observations of planets, owing to our lack of our knowledge
about observational efficiency, it is difficult to analysis the
distribution function of the parameters of the planet. There have been some
efforts to develop a fully deterministic strategy through automated
searching system for exoplanets \cite{matrin2010}. Having such a
system to cover all the known microlensing candidates will enable
us to obtain a correct statistical distribution of the
parameters of the planet. The other important channel for exoplanet
observation is that of high-magnification events \cite{safizadeh}. Almost all
the very high magnification events can be flagged by
microlensing surveys and follow-up telescopes monitor them with a
high sampling rate and better photometric precision. Unlike the
case for low-magnification events, for these events, the detection efficiency
function is almost known and statistical analysis in the
parameter space can be applied \cite{gould2010}. One of the main
problems with the binary lenses is the $d\leftrightarrow d^{-1}$
degeneracy problem, whereby we can have almost the same light curve
for the close and wide binary lenses.

In the caustic classification of the binary systems we have three
type of topologies for the caustics and corresponding critical lines
\cite{sch86}, the so-called the "close", "wide" and "intermediate" or
resonance binaries. Fig. (\ref{cat}) shows these three categories
of caustic lines for three values of planet to the star
mass ratios. To study the wave optics feature during the caustic
crossing, we generate synthetic light curves for the three
categories and compare the results with the geometric optics features. Our
aim is to study the wave optics signals in the resonance and high-magnification channels.

In order to quantify the wave optics feature in the light curve, we
use the $\chi^2$ difference from the best fit of the wave optics and
geometric optics. Assuming $\sigma_i$ as the error bar for each data
point, $\mu_{i}^{(g)}$ as the magnification in the geometric optics
and $\mu_{i}^{(w)}$ as the magnification in the wave optics, the
difference between the $\chi^2$s is given by
\begin{eqnarray}
\Delta\chi^2 &=& \chi^2_{g} - \chi^2_{w} \nonumber \\
&=& \sum_{i=1}^N \frac{1}{\sigma_i^2}(\mu_{i}^{(w)}- \mu_{i}^{(g)})(2\mu_i^{(exp)} - \mu_{i}^{(w)} - \mu_{i}^{(g)}).
\label{dchi2}
\end{eqnarray}
Having a threshold for $\Delta\chi^2$, we can distinguish the wave
optics light curve from the for the geometric optics. An important element
in equation (\ref{dchi2}) is the estimation of the photometric error
which depends on the source flux, integration time and the size of
the radio telescope. Amongst the various sources, red giants and
super-giants can emit electromagnetic waves at longer
wavelengths. Radio-loud Quasars at the cosmological scales are also
bright radio sources.
% where caustic crossing features can be
%observed in the cosmological scales.

Detailed studies on radio sources are performed for single-lens wave
optics microlensing in \cite{heyl11a}. In what follows we adapt that
classification. For the red giants, a closer star such as Arcturus
\cite{Perry} emits at the wavelengths of 2cm and 6cm with 0.68 mJy
and 0.28 mJy, respectively \cite{Drake}. The spectrum of this type
of star is given by
\begin{equation}\label{giants}
f_{\nu}\simeq
24~(\frac{\nu}{GHz})^{0.8}(\frac{kpc}{D_{s}})^{2}~\rm{nJy}.
\end{equation}
Another class is that of low-mass late-type stars as asymptotic giants.
For example, Mira is an example of this class and its spectrum is
given by \cite{Perry,asym1}
\begin{equation}\label{mira}
f_{\nu}\simeq 72~(\frac{\nu}{GHz})^{2}
(\frac{kpc}{D_{s}})^{2}~\rm{nJy}.
\end{equation}
Finally super-giants have strong radio emission. Betelgeuse is a
 red super giants located at a distance of $197$ pc
\cite{Newell,Harper}. The spectrum of this star normalized to the
kiloparsec distance is
\begin{equation}\label{supergiants}
f_{\nu}\simeq 9.3
~(\frac{\nu}{GHz})^{1.32}(\frac{kpc}{D_{s}})^{2}~\mu \rm{Jy}.
\end{equation}

The radii of super-giants are of the order of a few astronomical
unit. Assuming these stars are in the Galactic center at $\sim 8 kpc$
distance from us, from equation (\ref{ls}), we can obtain coherent
images for close-images on the lens plane. As mentioned above,
at the caustic crossing, only a small part of the source contributes
to the production of coherent close-pair images.

In equation (\ref{dchi2}), we need an estimation for the photometric
error bar. For the SKA project the noise corresponds to Nyquist
sampling is $0.27$ Jy \cite{Sci07}. This sampling is defined
such that the integration time multiplied by the band width, is
equal to one (i.e. $\Delta \nu ~\Delta\tau = 1$). Because noise
decreases with the square root of time and the band width as $
1/\sqrt{\Delta \nu \Delta\tau}$, we can write the noise in terms of
these two parameters as follows:
\begin{equation}
\Delta f = 0.14 (\frac{\Delta \nu}{GHz})^{-1/2}
(\frac{\Delta\tau}{1hr})^{-1/2} \mu \rm{Jy}.
\end{equation}

We use the six parameters of the binary microlensing system to simulate
the light curves. Also the source star is assumed to radiate at
longer wavelengths. We compare the simulated data in the geometric
optics case with the wave optics case and use the criterion of
$\Delta\chi^2>5$ between the two theoretical light curves. For an
ensemble of light curves, we identify caustic lines in three
categories of binaries, as shown in Fig. (\ref{cat}). Those light
curves with caustic crossing satisfy the criterion for the wave
optics are identified in the figure. Our analysis shows that wave
optics feature is sensitive to the specific parts of the caustic lines
of a binary lens.

\begin{figure}
\begin{center}
\psfig{file=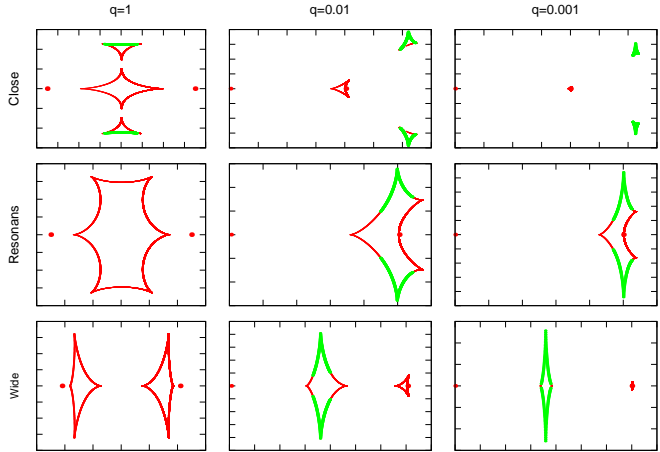,angle=270,width=9.cm,clip=}\caption{Caustic
lines for three categories of wide, close and intermediate
separation of two lenses with three different values of $q$. The
classification for the separation of the lenses is based on three
topology of the caustic lines (Schneider \& Wei$\ss$ 1986). The
thick-green lines of the caustic produces light curves with
$\Delta\chi^2>5$ in equation (\ref{dchi2}).  The binary lenses are
indicated by two spots and scales in the figures are given in terms
of distance between the lenses.} \label{cat}
\end{center}
\end{figure}

As we noted before, there are two main channels for the exoplanet
observations. In Figure (\ref{cat}) we identify area of caustic
lines in each channel with the wave optics feature. Having a small
$q$, wide and close binaries can produce almost the same geometric
optics light curves. According to Figure (\ref{cat}), there is no
strong wave optics effect in the light curve of the high
magnification events. On the other hand for the intermediate regime
(resonance) a larger area of the caustic lines is suitable for the
wave optics effect. In this case, we will have a combination of
close and wide images on the lens plane.

We perform a Monte-Carlo simulation to generate an ensemble of light
curves and study the wave optics effect in the simulated light
curves. Fig. (\ref{exper}) shows a sample of light curve in this
simulation with the cadence rate of $45$ min and signal-to-noise
ratio of ${S}/{N}=7$.
\begin{figure}
\begin{center}
\psfig{file=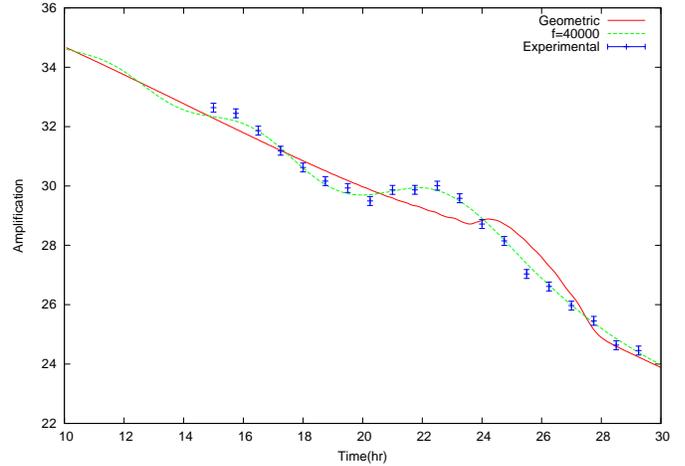,angle=270,width=9.cm,clip=} \caption{Simulation
of the microlensing light curve in a binary lens. The parameters of
the lens is taken as the light curve in Figure (\ref{caus-fig}). The
data points are simulated according to the noise in SKA. Here we
have a signal-to-noise of 7 with 45 minute integration time. The
observation is done in $10$ GHz. The solid line represents the best
fit in the geometric optics and the dashed line represents the best
fit of the light curve including the the wave optics effect.}
\label{exper}
\end{center}
\end{figure}
The relevant parameters of the wave optics from equation
(\ref{Ampli-Airy}) are the wavelength in the Airy function, $Y_0$
and the maximum magnification of the light curve, $\mu_{max}$. These
parameters depend on $f$ and $\phi_{222}$. For an ensemble of light
curves distributed  uniformly in parameter space, we calculate the
$\chi^2$ difference between the wave optics and the geometric optics
from equation (\ref{dchi2}). In order to study the sensitivity of
discriminating parameter, $\Delta\chi^2$, in terms of $f$ and
$\phi_{222}$, we identify area of parameter space that satisfies
$\Delta\chi^2>10$ where the source sizes are chosen $\rho =0.001,
0.005$ and $0.002$, see Fig. (\ref{fig:dchi2}). Here the upper
part of these curves do not satisfy our criterion and are excluded.
Having smaller $f$ means longer wavelength for the observation. On
the other hand, $\phi_{222}$ relates to the maximum magnification and
size of modes from the wave optics. By measuring how the transit time-scale
of the fringes changes with respect to the observer $\Delta
\tau$, we can determine $Y_0(f,\phi^{(0)}_{222}) = \Delta \tau/t_E$.
However, $\mu_{max}(\phi_{222})$ can be measured directly
from the light curve.
\begin{figure}
\begin{center}
\psfig{file=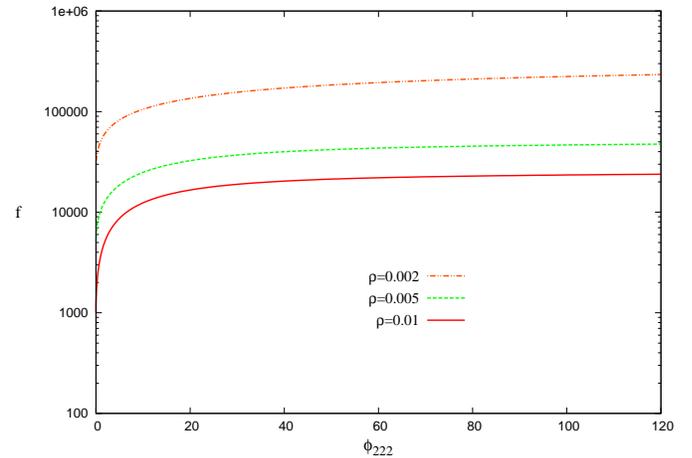,angle=270,width=9.cm,clip=} \caption{Wave optics features
are excluded in the area above the curves in the parameter space of $f$ and $\phi_{222}$ according to criterion
$\Delta\chi^2>10$. The size of the source star normalized to the
Einstein radius is mentioned in the legend of Figure. Smaller stars
are favorable for the observation of the wave optics features in
microlensing.} \label{fig:dchi2}
\end{center}
\end{figure}

The physical parameters involved in the wave optics light curves are
the overall mass of system $M_t$, the mass ratio $q$ and trajectory of the
source with respect to the lens. In contrast, from
observations in the geometric optics case, we can find six parameters of the
light curve with a degree of degeneracy. Having extra information from
the wave optics will constrain  $M_t$ and $q$ parameters and
subsequently we can identify the parameters of planet with better
accuracy.

We now want to look at the sensitivity of the wave optics signals in
terms of the physical parameters of the binary system, $q$ and $d$.
From the Monte-Carlo simulation, we select a fraction of events that
satisfy the condition $\Delta\chi^2 > 10$. According to Fig.
(\ref{percent}), a suitable area of the parameter space for the wave
optics features is in the resonance area where the separation
between the lens and the planet is of the order of Einstein radius.
This result is compatible with our preliminary analysis for the
sensitivity of wave optics signal in terms of the parameter space in
Fig. (\ref{cat}). In order to estimate the overall number of
microlensing events with the wave optics signals, we should multiply
the efficiency function with the real distribution of binary lenses in
terms of $q$ and $d$.
\begin{figure}
\begin{center}
\psfig{file=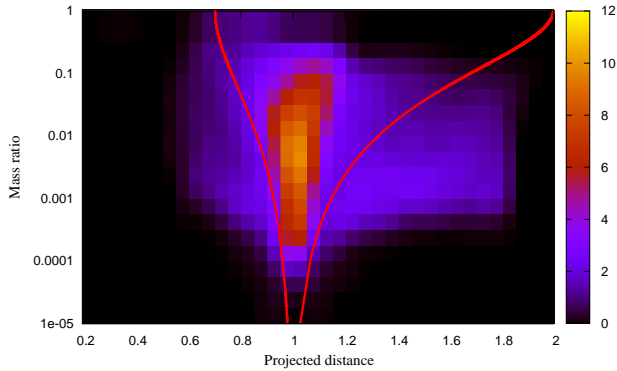,angle=270,width=9.cm,clip=} \caption{Efficiency
function in terms of $d$ and $q$, comparing the wave optics features
with the geometric optics. The criterion for the wave optics light
curve from equation (\ref{dchi2}) is $\Delta\chi^2>10$. The
simulation is done for the source star with the finite size of
$\rho=0.005$ and observation at $\nu = 10GHz$. The six parameters of
the lens are generated with the uniform distribution. Red lines
separate the parameter space according to the definition of the
wide, close and intermediate
 binaries (Erdl \& Schneider 1993).}
\label{percent}
\end{center}
\end{figure}

Finally, we want to extract physical information from a typical wave
optics light curve, assuming that we have observational data at both
visual and radio wavelengths. From the data of geometric optics,
by fitting to the light curve we can extract $q$, $d$ and the trajectory
of the source. On the other hand, from the wave optics our relevant
variables are $f$ and $\phi_{222}\sim 1/\sqrt{q}$. Measurement of these
two parameters provides directly the value of $q$ and the overall
mass of the lenses. Assuming a set of simulated data points of the
light curve, let us extract the observable parameters from this
light curve. We fit the simulated data in Fig. (\ref{sec-exp})
with the theoretical wave optics light curve. Here the theoretical
value of $f$ is $49475$, assuming that observations are performed at $\nu =
10$ GHz, and from the likelihood function we obtain $f =
51400_{-1600}^{+1635}$. Using $k$ from the definition of $f$, we can
extract the overall mass of a binary system. We can also extract the
mass radio of the planet to the parent star from the wave optics
light curve.

\begin{figure}
\begin{center}
\psfig{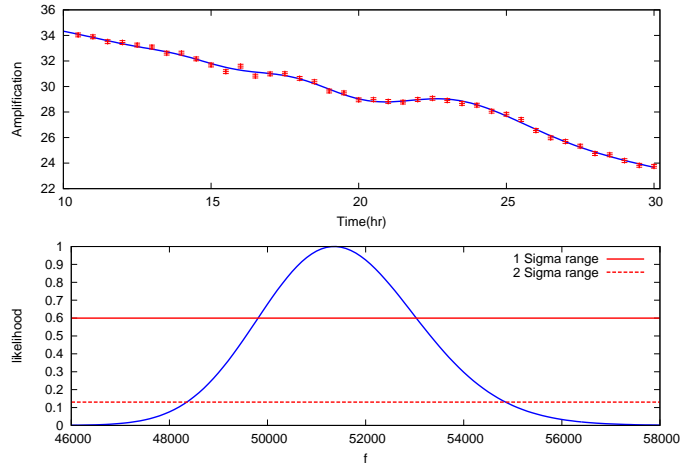} \caption{The synthetic
light curve in the radio wave lengths is generated with the cadence
of $30$ min and error bar of $0.15$ and fitted with the wave optics
theoretical light curve. The initial value of $f$ is taken $49475$.
We use $\nu=10GHz$ in the simulated light curve for the observation.
The likelihood function shows the best fit with $1\sigma$ and
$2\sigma$ level of confidences. For $1\sigma$, we have $f =
51400_{-1600}^{+1635}$, provides $6\%$ uncertainty in mass
measurement.} \label{sec-exp}
\end{center}
\end{figure}

\section{Conclusion}
\label{conc} In this work we studied the wave optics effect of
gravitational microlensing by a binary lens composed of a lens star
and a companion planet. The lensing effect of a planet during the
caustic crossing produces close images, a suitable configuration
for the images on the lens plane to generate diffraction pattern on
the observer plane. This effect is an example of Young's double-slit
experiment in the astronomical scales. We derived the wave optics
features of a binary lens, showing that it depends only on the third-order
derivatives of the Fermat potential $\phi_{222}$ and $f = 2k
R_s$.

We take red giants and super-giants as the source stars of
gravitational microlensing toward the Galactic bulge, as there is a
natural selection bias for observing this type of source stars. We
suggested using SKA future project for the observation of the wave
optics signals in the light curve. In this observational program,
radio observatories accompany the microlensing follow-up telescopes
in the visual bands. These two observations at long and short
wave lengths can provide a complimentary program for studying binary
microlensing events to break the degeneracy of binary systems.
We discussed the problem of spatial coherency of sources in
binary lensing and showed that only the part of the source that crosses
the caustic line contributes to the formation of close-pair images.
While an extended giant star may have no spatial coherency, the spatial coherency condition holds for the
small part of the source crossing the caustic line.

We studied the observability of the wave optics parameters in a
Monte Carlo simulation by fitting the simulated microlensing light
curves with the theoretical wave optics light curve. Out of the two
channels for the detection of exoplanets by microlensing, namly (i) the high-magnification
channel and (ii) the resonance channel, we showed that the
wave optics observation is in favour of resonance binary microlensing events.  The extra
information from the wave optics light curve enable us to solve for the
lens parameters with better accuracy. Our analysis has shown that the use
of radio telescopes for observations of planetary microlensing
events will open a new window for studies of exoplanet .

\section*{Acknowledgments}
AM thanks Sharif university of Technology providing high performance
computational facilities. We thank Avery Broderick, Vahid Karimipour
and Mir Abbas Jalali for providing useful comments and improving
text of paper. We also would like to thank anonymous referee for
valuable comments. This research was supported by Perimeter
Institute for Theoretical Physics and the John Templeton Foundation.
Research at Perimeter Institute was supported by the Government of
Canada through Industry Canada and by the Province of Ontario
through the Ministry of Economic Development and Innovation.

\appendix
\section{Coherency in Wave optics: Gravitational lensing}
\label{appA}In this appendix we adapt the wave optics
notations used in Schneider et al. (1992). Starting from the
amplitude of electromagnetic waves on the lens plane, we can obtain
the amplitude on the observer plane from the superposition principle
as
\begin{equation}
V = \int e^{if\phi(x,y)} d^2x. \label{a1}
\end{equation}
We perform a Taylor expansion of the Fermat potential around images and
diagonalize the second-order derivatives of the potential. The
amplitude of the electromagnetic waves on the observer plane is
\begin{equation}
V = \int e^{if[\phi^{(0)}+ \frac{1}{2}(\phi_{11}^{(0)} x_1^2 +
\phi_{22}^{(0)}x_2^2)]} d^2x. \label{a2}
\end{equation}
Integrating from equation (\ref{a2}), we find
\begin{equation}
V = \frac{2\pi i}{f}\frac{1}{\sqrt{\det|J|}}e^{if(\phi^{(0)}
-n_i\pi/2)}, \label{a3}
\end{equation}
where $\det|J|$ is the determinant of $\phi_{ij}^{(0)}$, which is
given in equation (\ref{jackobian}). Here $n$ refers to the type of
images and can be equal to $ n = 0, 1, 2$, depending on the number of
focal points transverse from the source to the observer
\cite{arnold}. Now if we have $N$ images from the lensing, the
overall amplitude is given by
\begin{equation}
V = \frac{2\pi i}{f}\sum_{i=1}^N\frac{e^{i(f\phi^{(0)}_i - n_i\pi/2)}}{\sqrt{|\det{J_i}|}}.
\end{equation}
We note that this equation is valid while the source is out of the
caustic lines (i.e. $\det{|J_i|}\neq 0)$. We assume a spectrum for
the source and replace $f$ with $2\omega R_s$, the overall amplitude
can be written as
\begin{equation}
V = {2\pi
i}\sum_{i=1}^N\frac{1}{\sqrt{|\det{J_i}|}}\int\frac{1}{f(\omega)}
e^{i(2R_s \omega\phi^{(0)}_i - n_i\pi/2)} g(\omega) d\omega.
\label{v}
\end{equation}
Finally, the overall magnification obtain from $\mu = V V^\star$.
After averaging over time, the magnification is given by
\begin{eqnarray}
\label{mut}
\mu &=& \sum_{i = 1}^N \mu_i  \\
&+& \sum_{i\neq j }
\frac{4\pi^2}{\sqrt{|\det{J_i}||\det{J_j}|}}\int\frac{
g(\omega)^2}{f(\omega)^2} e^{i[f(\omega)(\phi_i^{(0)} -
\phi_j^{(0)}) - \frac{\pi}{2}(n_i-n_j)]} d\omega. \nonumber
\end{eqnarray}
Here, time integration is done over the oscillating terms of
$e^{i\omega_it}$ and for the cross terms the result of integration
is a Dirac-Delta function. The phase term of
$f(\omega)(\phi_{1}^{(0)}-\phi_{2}^{(0)})$ in the integrand depends
on the phase difference between the $i$th and $j$th sources. For the
case that this phase is larger than the coherent time of the source,
the result of this integral is zero and only the first term of
equation (\ref{mut}) is non-zero, representing the geometric optics
contribute to the magnification. For the simple case of two images,
$N = 2$, let us assume a Gaussian spectrum for $g(\omega)$ with a
width given by $\Delta\omega$, the magnification for this simple
case obtain as follows:
\begin{equation}
\label{wom}
\mu=\mu_{1}+\mu_{2}+2e^{-(\frac{\Delta t}{\tau})^{2}}\sqrt{\mu_{1}\mu_{2}}\cos[\omega_{0}\Delta t-\frac{\pi}{2}(n_{1}-n_{2})],
\end{equation}
where we replaced the width of the spectrum with the coherent time as
$\Delta \omega=\frac{1}{\tau}$. For $\Delta t \ll \tau$, we have an
oscillating mode that reveals the wave optics feature from the
superposition of the waves. In contrast, for $\Delta t \gg \tau$ the
exponential suppresses the oscillating term and we will have a
geometric term for the magnification.

Now let us consider an extended source, in which each incoherent point
on the source contributes to the amplitude of the electromagnetic
waves on the observer plane. Hence for this case we can write
equation (\ref{v}) for each point of the source, assigning it by $V(s)$.
The overall amplitude can be written as

\begin{eqnarray}
\label{fs}
&&|V|^2 = \frac{4\pi^2}{S^2}\int\int \sum_{i,j=1}^N\frac{ds~ds'}{\sqrt{|\det{J_i(s)}| |\det{J_j(s')}|}}\times \\
 &&\int\int\frac{<g(\omega)g(\omega')>}{f(\omega)f(\omega')} e^{i[2R_s (\omega\phi^{(0)}_i(s)-\omega'\phi^{(0)}_j(s'))  - (n_i-n_j)\pi/2]}  d\omega  d\omega',\nonumber
\end{eqnarray}
where $S$ is the area of the source and averaging is performed over
time, hence $<g(\omega)g(\omega')> = g(\omega)^2\delta(\omega -
\omega')$. Because the differential elements on the source are
spatially uncorrelated, the cross terms in equation (\ref{fs}) will
 cancel, and only light rays propagating from the individual
elements of the source contribute in this summation. Mathematically
we can write
$$<\phi_i^{(0)}(s)\phi^{(0)}_j(s')> \sim \delta(s-s').$$ Hence
equation (\ref{fs}) simplifies to
\begin{equation}
\mu = \frac{1}{S} \int ds\int \frac{g(\omega)^2}{f(\omega)^2}
\sum_{i,j=1}^N\frac{e^{i[f(\omega)(\phi^{(0)}_i(s)-\phi^{(0)}_j(s))-
(n_i-n_j)\pi/2]} }{\sqrt{|\det{J_i(s)}| |\det{J_j(s)}|}} d\omega;
\end{equation}
or, in other words, the overall magnification can be written as
\begin{equation}
\mu = \frac{1}{S}\int\mu(s) ds. \label{fss}
\end{equation}

\end{document}